# Advanced microwave photonic waveform editing: enabling the evolution of radar systems into joint radar and spectrum sensing systems

Chi Jiang, Taixia Shi, Dingding Liang, Lei Gao, Chulun Lin, and Yang Chen[*]

Shanghai Key Laboratory of Multidimensional Information Processing, School of Communication and Electronic Engineering, East China Normal University, Shanghai 200241, China
[*]Correspondence to: ychen@ce.ecnu.edu.cn

**ABSTRACT**
In response to the urgent demand for the development of future radar application platforms from single radar functionality towards integrated multi-functional systems, we show an advanced microwave photonic waveform editing method that enables the editing of arbitrary radar waveforms, equipping them with the capability to perform spectrum sensing. This, in turn, expands single-function radar systems into joint radar and spectrum sensing systems. We theoretically define and calculate the accumulation function of an arbitrary waveform after passing through a specific dispersive medium, and utilize this accumulation function to further design a corresponding binary sequence for editing the waveform. After editing, the accumulation function of the edited waveform approximates that of a linearly frequency-modulated signal matching the specific dispersive medium. Thus, the edited waveform can be compressed into a narrow pulse after passing through the dispersive medium, realizing the frequency-to-time mapping for achieving frequency measurement or time–frequency analysis. The concept is verified by a simulation and an experiment. Using a dispersion compensating fiber with a total dispersion of −6817 ps/nm, arbitrary waveforms, including a 7-bit Barker phase-coded waveform, a linearly frequency-modulated waveform, a nonlinearly frequency-modulated waveform, and a waveform with an "E" time–frequency diagram, are edited and further used for microwave frequency measurement and time–frequency analysis in an ultra-wide bandwidth of 36.8 GHz. The temporal resolution and frequency resolution are 2 ns and 0.86 GHz, respectively.

## 1. Introduction
As a technology capable of realizing accurate detection of multi-dimensional information on targets, radar can meet the application requirements of multiple fields, such as detecting obstacles and other vehicles to ensure the safety in autonomous driving vehicles [[1], [2]]，reconnoitering enemy targets and guiding weapons to destroy them in modern war [[3]]. Accurate detection requires radar signals with a larger bandwidth, and microwave photonic radar can easily generate such radar signals [[4]]–[[7]]. Besides the single function of detecting vehicles in the field of autonomous driving, it is also very important to enable the radar system to possess the ability to cognitively sense and utilize the spectrum, which can help vehicles reasonably plan the operating frequency bands of their various electromagnetic functions [[8], [9]]. In electronic warfare [[10]], the integration of radar and spectrum sensing functions is crucial for enhancing cognitive capabilities and facilitating further intelligence decision-making. Therefore, in the future, the evolution of radar application platforms from single-function radars toward multi-functional integration is a widely recognized development trend. Besides, as spectrum resources become increasingly scarce and the electromagnetic environment becomes more complex, radar systems also need to develop the capacity to understand their surrounding spectrum environment and enhance their anti-jamming ability and survivability [[11]]. The joint radar and spectrum sensing system is capable of simultaneously detecting the target and analyzing the spectrum usage of the surrounding electromagnetic environment, providing the foundation for the intelligence and self-adaptive capabilities of the radar system.

For future broadband application scenarios, as well as scenarios with extremely high real-time requirements, the core functions of the spectrum sensing system, namely frequency measurement and time–frequency analysis, need to possess ultra-high measurement speed and ultra-large measurement bandwidth. Traditional electronic technology is limited by inherent electronic bottlenecks and has poor real-time performance in broadband spectrum sensing. Microwave photonic spectrum sensing systems use photonic technology to perform fast measurements within a broad bandwidth and are considered a feasible solution to address the contradiction between bandwidth and real-time performance in traditional electronic solutions. Microwave photonic spectrum sensing methods can be implemented by frequency-to-power mapping (FTPM) [[12], [13]] and frequency-to-time mapping (FTTM) [[14]]–[[26]]. The FTPM-based methods, through the amplitude comparison function, map the microwave frequencies to optical or electrical signal power, which is simple and efficient. However, most FTPM-based methods can only be used to measure single-tone signals. The FTTM-based methods are more suitable for measuring multi-tone and even more complex signals, which are commonly encountered in practical applications. The FTTM-based methods can be mainly divided into two categories based on the specific principles: methods via optical frequency-sweeping and filtering [[14]]–[[21]], and methods via dispersion [[22]]–[[26]].

The former, via optical frequency-sweeping and filtering, adds a time window to the signal under test (SUT) using a short-period optical linearly frequency-modulated (LFM) signal, and a narrowband optical filter is used to implement FTTM in each sweep period after the optical LFM signal is modulated by the SUT. However, there is a contradiction between the real-time performance, frequency resolution, and measurement bandwidth of this kind of method: in the case where the measurement bandwidth is determined, high real-time performance requires an ultra-high sweep chirp rate; the filter bandwidth to obtain the optimal frequency resolution at a high sweep chirp rate needs to be increased rather than decreased, which is the conclusion we analyzed and verified in [[17]]. Therefore, if high real-time performance is required, such as a sweep chirp rate above GHz/ns, the required filter bandwidth is too large, making it difficult to achieve the measurement under this principle. This is also why methods based on optical frequency-sweeping and filtering generally have a sweep chirp rate in the GHz/μs level. The FTTM of this kind of method can also be implemented using electrical filters [[18]]–[[20]], but the above limitation also exists. Therefore, achieving an analysis bandwidth above GHz within 1 ns is beyond the capability of this kind of method. An improved method based on channelization has been reported as a potential solution [[21]], but it is relatively complicated.

The latter achieves FTTM by compressing optical signals with quadratic phase in the temporal domain through dispersion [[22]]–[[25]]. Here, optical signals with quadratic phase in the temporal domain can be realized either by broadening narrow optical pulses via dispersion [[23]] or directly modulating an LFM signal on an optical carrier [[24], [25]]. Relevant reports recently have expanded the analysis bandwidth of this kind of method to tens of GHz or even greater [[24]]–[[26]], while keeping a temporal resolution of several nanoseconds. The advantage in temporal resolution makes this kind of method more suitable for applications that require extremely high real-time performance compared to the method using optical frequency-sweeping and filtering.

As discussed above, LFM signals are utilized in both of the aforementioned two types of spectrum sensing methods. In fact, LFM signals are first and foremost widely used as one of the most common radar signals. Due to the widespread application of LFM signals in modern radar systems, when radar and spectrum sensing systems are to be integrated into one system through shared hardware and signals, some researchers have proposed microwave photonic joint radar and spectrum sensing systems based on LFM signals [[27]]–[[30]]. In these schemes, LFM signals are utilized to realize radar function, while their replicas in the optical domain are employed for spectrum sensing based on the principle of optical frequency-sweeping and filtering. Nevertheless, for radars other than LFM radars, such as phase-coded radars, there are currently no relevant reports on microwave photonic joint radar and spectrum sensing. This is mainly because other kinds of radar signals, beyond LFM radar signals, are difficult to use for spectrum sensing, thereby hindering the true hardware and signal sharing. Therefore, equipping other radar systems that utilize signals beyond LFM signals with the capability to be expanded into a joint radar and spectrum sensing system has become crucial for developing future radar application platforms from single radar functionality towards integrated multi-functional systems. However, this issue has not been well considered and resolved to date.

In this paper, we propose an advanced microwave photonic waveform editing method. This method aims to edit arbitrary radar waveforms and equip them with the capability to perform spectrum sensing, ultimately expanding single-function radar systems into joint radar and spectrum sensing systems. We theoretically define and calculate an accumulation function of an arbitrary waveform after passing through a specific dispersive medium, and utilize this accumulation function to further design a corresponding binary sequence for editing the arbitrary waveform. After editing, the accumulation function of the edited waveform approximates that of an LFM signal matching the specific dispersive medium. Thus, the edited waveform can be compressed into a narrow pulse after passing through the dispersion medium, realizing the FTTM for achieving frequency measurement or time–frequency analysis. Besides, the binary sequence can be generated by a signal generator equipped with a 1-bit DAC, which is much less costly compared to directly generating an LFM signal for spectrum sensing. The concept is verified by a simulation and an experiment. Using a dispersion compensating fiber (DCF) with a total dispersion of −6817 ps/nm, arbitrary waveforms, including a 7-bit Barker phase-coded waveform, an LFM waveform, a nonlinearly frequency-modulated (NLFM) waveform, and a waveform with an "E" time–frequency diagram, are edited and further used for microwave frequency measurement and time–frequency analysis in an ultra-wide bandwidth of 36.8 GHz. The temporal resolution and frequency resolution are 2 ns and 0.86 GHz, respectively.

**2. Principle of advanced microwave photonic waveform editing**

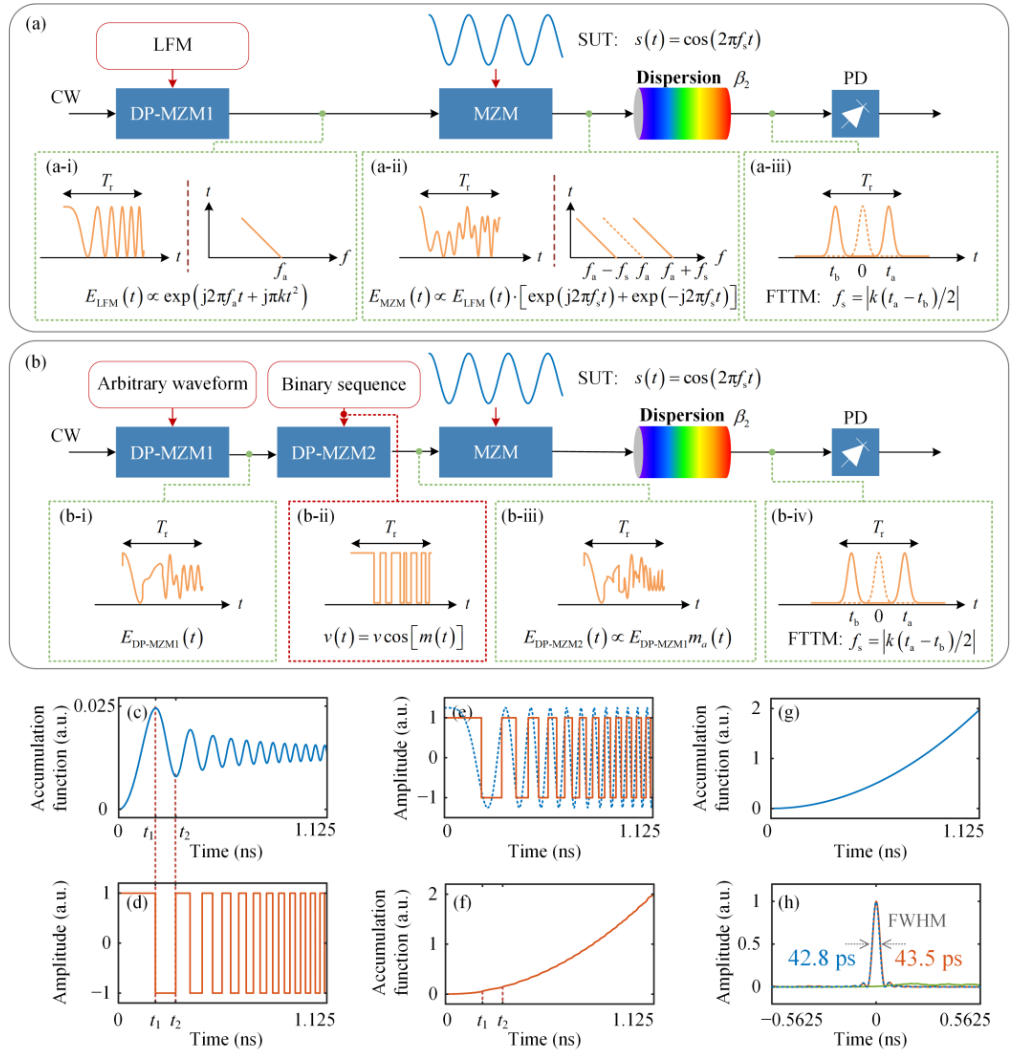

Fig. 1. (a) Schematic of the typical spectrum sensing methods based on dispersion and FTTM. (b) Schematic of the spectrum sensing method based on microwave photonic waveform editing, dispersion, and FTTM. (c) Accumulation function when a CW light wave directly passes through the DCF. (d) Temporal waveform of the designed binary sequence for the CW light wave. (e) Temporal waveform of an LFM signal (blue dashed line) and the binary sequence for a CW light wave (orange solid line) with a temporal duration of 1.125 ns. (f) Accumulation function when the CW light wave edited by the binary sequence passes through the DCF. (g) Accumulation function when an optical LFM signal passes through the DCF. (h) Optical pulses when an LFM signal is directly compressed (blue dashed line), when a CW light wave is edited and then compressed (orange solid line), and when the CW light wave is not edited (green solid line).

Advanced microwave photonic waveform editing is the key to evolving a radar system into a joint radar and spectrum sensing system. To reveal how to edit the radar waveforms, we first analyze the signal evolution process of the typical frequency measurement and time–frequency analysis methods based on dispersion and FTTM, as shown in Fig. 1(a). A CW light wave is modulated by an electrical LFM signal and converted into an optical LFM signal at a dual-parallel Mach–Zehnder modulator (DP-MZM1) via carrier-suppressed single-sideband (CS-SSB) modulation. The real part of the complex envelope of the optical LFM signal and its corresponding time–frequency diagram are shown in Fig. 1(a-i). Then, the optical LFM signal is further modulated by the SUT at a Mach–Zehnder modulator (MZM), which is biased as a carrier-suppressed double-sideband (CS-DSB) modulator. For ease of analysis, we assume that the SUT is a single-tone signal with a frequency of $f_s$, so the complex envelope of the two 1st-order optical sidebands from the MZM can be expressed as

$$E_{\text{MZM}}(t) \propto \exp\left[j2\pi(f_a + f_s)t + j\pi kt^2\right] + \exp\left[j2\pi(f_a - f_s)t + j\pi kt^2\right] \quad (1)$$

where $f_a$ and $k$ are the start frequency and chirp rate of the optical LFM signal from DP-MZM1. The real part of the complex envelope of the optical signal from the MZM and its corresponding time–frequency diagram are shown in Fig. 1(a-ii). The dashed line in Fig. 1(a-ii) indicates that the original optical LFM carrier is suppressed due to the CS-DSB modulation.

After the optical signal in Eq. (1) passes through a DCF that functions as a dispersion medium, in one sweep period $T_r$, the output optical signal can be expressed as

$$\begin{aligned}
E_{\text{output}}(t) &= E_{\text{MZM}}(t) * \exp\left(\frac{jt^2}{2\beta_2 L}\right) \\
&= \int_0^{T_r} E_{\text{MZM}}(\tau) \exp\left[\frac{j(t-\tau)^2}{2\beta_2 L}\right] d\tau \\
&= \int_0^{T_r} \left\{\exp\left[j2\pi(f_a + f_s)\tau + j\pi k\tau^2\right] + \exp\left[j2\pi(f_a - f_s)\tau + j\pi k\tau^2\right]\right\} \\
&\quad \times \exp\left[\frac{j(t-\tau)^2}{2\beta_2 L}\right] d\tau
\end{aligned} \quad (2)$$

where $\beta_2 = -\lambda^2 D/(2\pi c)$ is the dispersion coefficient, $L$ is the length of the DCF, $\lambda$ is the center wavelength of the input optical signal, $D$ is the dispersion parameter of the DCF, and $c$ is the light speed in a vacuum, and the symbol "*" means convolution. When the chirp rate of the optical LFM signal matches the total dispersion of the DCF, that is $k = -1/(2\pi\beta_2 L)$, Eq. (2) can be further derived and simplified as

$$\begin{aligned}E_{\text{output}}(t) &= \exp(-j\pi kt^2)\exp\left[j\pi(f_a+f_s+kt)T_r\right]\text{Sa}\left[\pi(f_a+f_s+kt)T_r\right]T_r \\ &+ \exp(-j\pi kt^2)\exp\left[j\pi(f_a-f_s+kt)T_r\right]\text{Sa}\left[\pi(f_a-f_s+kt)T_r\right]T_r\end{aligned} \quad (3)$$

As can be seen, the optical sidebands of the single-tone SUT are compressed into two symmetric pulses, as shown in Fig. 1(a-iii). The time at which two pulse peaks appear can be determined by

$$\begin{cases} t_a = -\dfrac{f_a+f_s}{k}, & \text{for } +1\text{st-order optical sideband of the SUT} \\ t_b = -\dfrac{f_a-f_s}{k}, & \text{for } -1\text{st-order optical sideband of the SUT} \end{cases} \quad (4)$$

As indicated by Eq. (4), the time at which the two pulses appear within one sweep period $T_r$ can represent the frequency of the SUT. Thus, the frequency of the SUT is mapped to the temporal domain, and by analyzing the pulse appearance time of one or more sweep periods, the frequency and time–frequency information of the SUT can be obtained. It is noted that in Fig. 1(a-iii), a dashed pulse is shown, which corresponds to the mapped pulse when no SUT is applied, i.e., $f_s = 0$. The dashed pulse is compressed at the moment $t = 0$ when the start frequency of the LFM signal is $f_a = 0$. As can be seen, with the increase in the frequency of the SUT, the two generated pulses will move farther away from this dashed pulse. To determine the appearance time of the two pulses, the optical pulses in Eq. (3) are converted to electrical pulses via a photodetector (PD) and captured by an oscilloscope (OSC). Within the analysis bandwidth, the frequency $f_s$ of SUT can be calculated according to the appearance time difference $\Delta t$ of the two pulses, which can be expressed as

$$f_s = \left|\frac{k(t_a-t_b)}{2}\right| = \left|\frac{k\Delta t}{2}\right|. \quad (5)$$

Since the SUT is loaded onto the optical LFM signal in a CS-DSB manner at the MZM, the aforementioned two symmetric pulses are generated. In fact, if a DP-MZM is used instead of the MZM to load the SUT onto the optical LFM signal in a CS-SSB manner, only one pulse will be generated. In this case, the analysis bandwidth of the system can be doubled [[25]]. Besides, since it is necessary to know the absolute position of the pulse within one sweep period in this case, synchronization is required between the signal sampling end and the LFM signal generation end [[25]].

The aforementioned pulse compression process shown in Fig. 1(a) requires that the optical signal input into the MZM be an optical LFM signal, which is readily achievable for LFM radars. This can be done by directly inputting the LFM signal of the LFM radar into DP-MZM1 or extracting a portion of the optical LFM signal directly from a microwave photonic LFM radar. However, for other types of waveforms, the aforementioned method is not feasible. Therefore, we aim to edit arbitrary waveforms in the optical domain so that they can acquire the pulse compression capability as demonstrated in the system shown in Fig. 1(a).

The principle of the advanced microwave photonic waveform editing and FTTM for frequency measurement and time–frequency analysis is shown in Fig. 1(b). Since the LFM signal in Fig. 1(a) is replaced by an arbitrary waveform as shown in Fig. 1(b-i), after the arbitrary waveform is CS-SSB modulated at DP-MZM1, another DP-MZM2 is employed in the system. DP-MZM2 is also biased as a CS-SSB modulator and driven by a binary sequence shown in Fig. 1(b-ii) to edit the waveform from DP-MZM1. Then the edited waveform from DP-MZM2 shown in Fig. 1(b-iii) is CS-DSB modulated by the SUT at the MZM, dispersed and compressed into optical pulses in the DCF, as shown in Fig. 1(b-iv), and then converted to electrical pulses after photodetection in the PD.

Then, how to design the binary sequence is the key issue. Here, we simplify the problem to the compression of an unmodulated optical signal from DP-MZM1 in a DCF. As discussed above, if the optical signal is an optical LFM signal and no SUT is applied, it should generate a pulse at $t = 0$ when $f_a = 0$ and the optical LFM signal matches the DCF. Let's consider this issue in this way: For an arbitrary optical waveform, within a period $T_r$, the contributions of the waveform at different time points to the intensity of the optical signal at $t = 0$ after processing by the DCF are different, i.e., the optical waveform at certain time points will increase the intensity, while others will decrease it. If we can calculate whether the impact of the optical waveform at different time points on the pulse generated at $t = 0$ is positive or negative, we can alter its influence on pulse generation by inverting the polarity of this waveform segment. For any waveform, if it can be compressed into a pulse at $t = 0$ after editing without applying the SUT, then after loading the SUT onto this edited waveform using the MZM in Fig. 1(b), it can similarly be compressed into two pulses that are symmetrical about $t = 0$.

It is assumed that no binary sequence and no SUT is applied, so the optical signal $E_{\text{DP-MZM1}}(t)$ directly passes through the DCF. Thus, the output signal from the DCF can be expressed as

$$E_{\text{output}}(t, T_r) = \int_0^{T_r} E_{\text{DP-MZM1}}(\tau) \exp\left[\frac{j(t-\tau)^2}{2\beta_2 L}\right] d\tau. \tag{6}$$

To calculate the contribution of the arbitrary waveform at different time points to the intensity of the optical signal at $t = 0$ after processing by the DCF, an accumulation function is introduced and calculated to guide the design of the binary sequence, which is expressed as

$$\text{AF}(t_r) = \left|E_{\text{output}}(t, T_r)\right|^2\bigg|_{t=0, T_r=t_r} = \left|\int_0^{t_r} E_{\text{DP-MZM1}}(\tau) \exp\left[\frac{j\tau^2}{2\beta_2 L}\right] d\tau\right|^2 \tag{7}$$

where $t_r \in [0, T_r]$. This is an integral with a variable upper limit, which reflects the variation in the influence that an arbitrary optical waveform, as it changes over time within one period, has on the signal intensity at $t = 0$ after processing by the DCF.

An example is given in Fig. 1(c) where a CW light wave is employed as the arbitrary waveform mentioned above. This can be considered a special case of single-frequency radar. Here, the total dispersion of the DCF is set to −6817 ps/nm. As can be seen from Fig. 1(c), the accumulation function oscillates with time. The waveform from time 0 to $t_1$ has a positive impact on pulse generation, while the waveform from time $t_1$ to $t_2$ has a negative impact, and so forth. Therefore, we can design a binary sequence based on the accumulation function: retaining the waveform sections with positive impact unchanged (i.e., multiplying by 1) and inverting the waveform sections with negative impact by multiplying them by −1. The designed binary sequence for the CW light wave is shown in Fig. 1(d), and the correspondence between the "+1" and "−1" in Fig. 1(d) and the oscillation relationship in Fig. 1(c) is very clear. After introducing the waveform editing, the accumulation function of the edited waveform can be expressed as

$$\begin{aligned}
\text{AF}(t_r) &= \left|\int_0^{t_r} E_{\text{DP-MZM2}}(\tau) \exp\left[\frac{j\tau^2}{2\beta_2 L}\right] d\tau\right|^2 \\
&= \left|\int_0^{t_r} E_{\text{DP-MZM1}}(\tau) m_a(\tau) \exp\left[\frac{j\tau^2}{2\beta_2 L}\right] d\tau\right|^2
\end{aligned} \tag{8}$$

where $m_a(\tau)$ is the term introduced by the binary sequence $v(t) = v\cos[m(t)]$ at DP-MZM2, $m(t)$ is "0" or "$\pi$", and $v$ is the amplitude of the binary sequence that only affects the power of the compressed pulse. Therefore, the portions of the optical waveform that positively contribute to generating the pulse at $t = 0$ are multiplied by 1, while the portions that negatively contribute are multiplied by −1. In the CS-SSB modulation, the phase term $m_a(\tau)$ is the analytic signal of $v(t)$, which can be obtained by applying the Hilbert transform to $v(t)$.

We provide an LFM signal with a temporal duration of 1.125 ns that matches the dispersion of the DCF, which means it can be compressed to a narrow pulse after being transmitted in the DCF. The LFM signal is presented in blue dashed line in Fig. 1(e), which is also the real part of the complex envelope of the optical waveform after the LFM signal is CS-SSB modulated at DP-MZM1. Additionally, Fig. 1(e) also displays in orange solid line the binary sequence with the same temporal duration used for editing the aforementioned CW light wave, which is also the real part of the complex envelope of the CW light wave after editing using the binary sequence. The amplitude of the two signals in Fig. 1(e) is different because we want to control them to have the same power. The accumulation function of the CW light wave after editing is shown in Fig. 1(f). Unlike the oscillating accumulation function in Fig. 1(c), the accumulation function here, although exhibiting slight fluctuations, exhibits an overall monotonically increasing trend. The accumulation function of the LFM waveform is given in Fig. 1(g). It is obvious to find that the accumulation function of the CW light wave after editing approximates that of the LFM waveform. Thus, the CW light wave after editing can be compressed into a narrow pulse after passing through the DCF. The blue dashed line in Fig. 1(h) is the temporal waveform of the optical pulse compressed from the optical LFM signal, and the orange solid line in Fig. 1(h) is the temporal waveform of the optical pulse compressed from the edited waveform of the CW light wave. For the originally difficult-to-compress CW light wave, after being edited using our proposed waveform editing method, it has shown a significant enhancement in the ability to be compressed into a pulse when compared to the waveform shown in in green solid line in Fig. 1(h), which is the waveform output from the DCF when the CW light wave directly passes through the DCF. In addition, because the dispersion of the DCF is fixed, the full-widths at half maximum (FWHMs) of the two pulses are almost identical. Therefore, when the signal period is determined, the analysis bandwidth, frequency resolution, and other parameters of the proposed system with advanced waveform editing are consistent with those of methods based on LFM signals with the same period [[25]]. The analysis bandwidth can be calculated by

$$B = \left| \frac{T_r}{2\pi \beta_2 L} \right|. \tag{9}$$

The frequency resolution of the system is associated with the FWHM of the pulse after FTTM. By squaring Eq. (3), the FWHM of the pulse is calculated to be around $|0.886/kT_r|$, so the theoretical frequency resolution can be expressed as

$$r_f = \frac{\text{FWHM}}{T_r} B \approx \frac{0.886}{T_r}. \tag{10}$$

## 3. Experimental setup and results

3.1 Experimental setup

The experimental setup of the proposed microwave photonic waveform editing method for spectrum sensing is shown in Fig. 2. A 13.5-dBm CW light wave centered at 1549.941 nm from an LD (ID Photonics, CoBriteDX1-1-C-H01-FA) is employed as the optical carrier and modulated by an arbitrary radar waveform at DP-MZM1 (Fujitsu FTM7961EX). DP-MZM1 is biased as a CS-SSB modulator using a modulator bias controller (PlugTech MBC-IQ-03). The output of DP-MZM1 is edited by the binary sequence at DP-MZM2, which is also biased as a CS-SSB modulator. The optical signal after editing from DP-MZM2 is amplified by an erbium-doped fiber amplifier (EDFA1, EDFA-C-PA-45-SM-M) to compensate for the insertion loss and conversion loss of the two modulators. The output of EDFA1 is modulated by the SUT at the MZM (Fujitsu FTM7938EZ). The MZM is biased as a CS-DSB modulator using a modulator bias controller (PlugTech MBC-NULL-03). EDFA2 (Amonics AEDFA-PA-35-B-FA) and EDFA3 (Max-Ray EDFA-PA-35-B) are utilized to perform pre-amplification and post-amplification, respectively, for the optical signal input into the DCF, aiming to compensate for the high insertion loss of the DCF (30 dB insertion loss and −6817 ps/nm dispersion). An optical bandpass filter (OBPF, WL Photonics WLTF-BA-U-1550-100) with a bandwidth of around 200 GHz and a center wavelength of around 1549.941 nm is used to filter out the out-of-band amplified spontaneous emission (ASE) noise. Then, the optical signal after OBPF is converted back to the electrical domain via a PD (u2tMPRV1331A). The electrical pulse generated from the PD is captured by an oscilloscope (OSC, LeCroy WaveMaster 820Zi-B, 80 GSa/s). Note that the arbitrary radar waveform, the binary sequence, and the SUT are all generated from a four-channel arbitrary waveform generator (AWG, Keysight M8195A, 64 GSa/s).

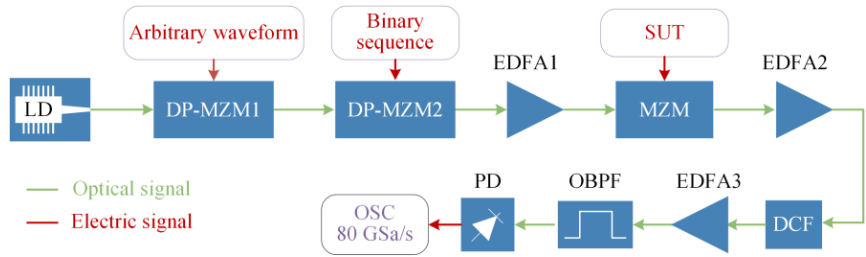

Fig. 2. Schematic diagram of the proposed microwave photonic waveform editing method for spectrum sensing. LD, laser diode; DP-MZM, dual-parallel Mach–Zehnder modulator; EDFA, erbium-doped fiber amplifier; MZM, Mach–Zehnder modulator; DCF, dispersion compensating fiber; OBPF, optical bandpass filter; PD, photodetector; OSC, oscilloscope.

3.2 Waveform Editing for CW Waveform

First, the waveform editing for single-frequency radar is demonstrated. In this case, the radar waveform is a CW signal. Here, a single-tone signal at 5 GHz is employed, which is sent to DP-MZM1. Thus, after CS-SSB modulation at DP-MZM1, the frequency-shifted optical carrier is still a CW light wave. A binary sequence is designed for the CW light wave according to the theory given in Section 2. Since the CW light wave is periodic, here we generate a 2-ns periodic signal, with the first 1.125 ns being a binary sequence for editing the waveform and the last 0.875 ns being zeros. Thus, the edited waveform can achieve a temporal resolution of 2 ns for spectrum sensing. For comparison, we also employ an LFM signal that possesses a frequency range from 0 to 20.7 GHz and a chirp rate of 18.4 GHz/ns. The LFM signal is matched to the DCF with the same structure as the binary sequence, that is 1.125-ns LFM signal followed by 0.875-ns zeros. The LFM signal does not need waveform editing. Figure 3(a) shows the temporal waveforms of the binary sequence in the experiment and simulation from 0 to 1.125 ns for waveform editing, as well as the corresponding LFM signal in the experiment and simulation. Comparing the temporal waveforms in Fig. 3(a), it can be found that the binary sequence and the LFM signal exhibit similar trends. However, the binary sequence can be generated by a signal generator equipped with a 1-bit DAC, which is less costly compared to the direct use of a multi-bit DAC for generating an LFM signal. Figure 3(b) shows the electrical spectra of the binary sequence and the LFM signal in the experiment. In practical applications, the signal bandwidth is generally limited. Therefore, when loading the signals into the AWG, we restrict the signal bandwidth to 20.7 GHz. Additionally, it should be noted that due to the frequency response of the AWG and the devices connecting the AWG to the modulator, the signal power is relatively lower at high frequencies. However, this does not affect the validation of the proposed method.

Based on the dispersion provided by the DCF and the signal temporal length, according to Eq. (9), the CW light wave with waveform editing, as well as the optical LFM signal, possesses an analysis bandwidth of 36.8 GHz when the duty cycle is taken into consideration [[25]]. Figure 3(c) shows the generated electrical pulses in a period when a single-tone signal at 10 GHz is employed as the SUT. The blue dashed line and the orange solid line are the pulses compressed from the optical LFM signal without waveform editing and the CW light wave with waveform editing, respectively. The FWHMs of the pulses are 49.0 ps and 46.9 ps, respectively, corresponding to frequency resolutions of about 0.90 GHz and 0.86 GHz. This is consistent with the theoretical value of 0.79 GHz obtained from Eq. (10).

After performing waveform editing on the CW signal in the optical domain, we further utilize the edited waveform to conduct frequency measurement and time–frequency analysis experiments. Figure 3(d) shows the captured electrical pulses in five individual measurements when single-tone signals at frequencies of 1, 5, 9, 13, and 17 GHz are employed as the SUT. The frequency measurement errors of 400 measurements at the five frequencies are shown in Fig. 3(e). The average measurement values of the five frequencies are 1.0568, 5.0381, 8.9338, 12.9456, and 16.9096 GHz, respectively, while the corresponding standard deviations are 113.0, 67.5, 83.9, 104.0, and 115.1 MHz, respectively. Figure 3(f) shows the time–frequency analysis result of an NLFM signal with a temporal length of 4 µs and a frequency range from 2 to 17 GHz using the edited waveform. As can be seen, the analysis bandwidth is from −18.4 to 18.4 GHz, and the time–frequency characteristic of the NLFM signal is well recovered. The symmetry of the analysis bandwidth and the analysis results is determined by the CS-DSB modulation at the MZM, and this has been explained in detail in Section 2.

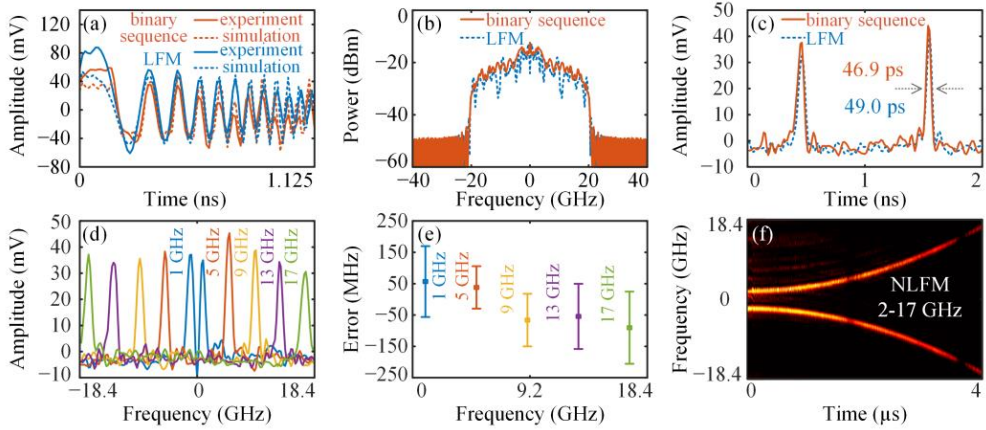

Fig. 3. (a) Temporal waveforms of the binary sequence and the LFM signal in the experiment and simulation. (b) Electrical spectra of the LFM signal and the binary sequence in the experiment. (c) Generated electrical pulses compressed from the optical LFM signal without waveform editing and from the CW light wave with waveform editing, when a single-tone signal at 10 GHz is employed as the SUT. (d) Generated electrical pulses when single-tone signals at frequencies of 1, 5, 9, 13, and 17 GHz are employed as the SUT. (e) Frequency measurement errors of the five single-tone signals. (f) Time–frequency diagram of an NLFM ranging from 2 to 17 GHz.

### 3.3 Waveform Editing for 7-bit Barker Phase-coded Waveform

Widely adopted in radar systems, the 7-bit Barker code serves as a prevalent code for phase-coded radar waveforms. It exhibits unique autocorrelation properties in radar signal processing, which is of great significance for enhancing the accuracy and reliability of radar detection. Then, we will edit the 7-bit Barker phase-coded waveform in the optical domain to endow it with the capability of spectrum sensing.

A 7-bit Barker phase-coded waveform with a temporal length of 1.09375 ns followed by 8.90625-ns zeros is employed in this experiment. The center frequency of the waveform is 7 GHz. First, the radar ranging function of the 7-bit Barker phase-coded waveform is verified. This validation is achieved equivalently by measuring differences in length. In this experiment, we transmit the radar waveform through two cables of different lengths, then collect the coupled waveform and process it through matched filtering. Due to the difference in transmission time of the signal through the two cables, two peaks will appear after matched filtering. The time difference between these two peaks, which represents the time difference in signal transmission through the two cables, can be used, along with the signal propagation speed in the cable, to calculate the length difference between the two cables via the following equation,

$$\Delta l = \Delta t_{radar} / \beta \tag{11}$$

where $\Delta t_{radar}$ denotes the time difference between the two peaks, $\Delta l$ represents the length difference of the two RF cables, and $\beta = 4.765$ ns/m is the time required for an electrical signal to transmit one meter in the cables.

In the experiment, the length difference is set to 0.25, 0.35, 0.65, 0.85, 1.35, and 1.85 m, respectively. The corresponding experimental results are shown in Fig. 4. As can be seen, under different cable length differences, narrow pulses with two distinct spacings are obtained after the 7-bit Barker phase-coded waveform passing through the two cables is match filtered. It can also be observed that in the same subfigure, the amplitudes of the two pulses are different, with the pulse corresponding to the longer cable having a lower amplitude. Furthermore, by comparing different subfigures, it can be seen that the pulse amplitude corresponding to the cable with a varying length in the experiment decreases continuously as the length of that cable increases. This is because a longer cable length introduces greater signal attenuation, thereby reducing the pulse amplitude after matched filtering. The time differences from Fig. 4(a) to Fig. 4(f) are 1.16, 1.62, 3.05, 4.00, 6.34, and 8.69 ns, respectively, so the corresponding cable length differences are 0.243, 0.340, 0.640, 0.839, 1.331, and 1.824 m. The measurement results are very close to the set values, with the maximum error being 2.6 cm.

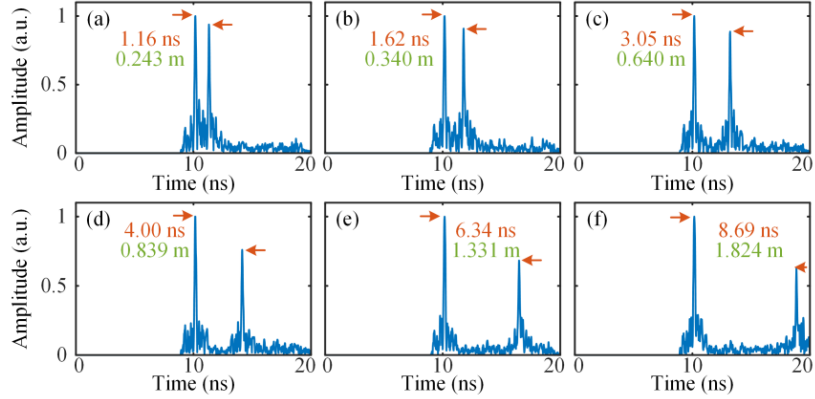

Fig. 4. Measurement results of the length difference between cables of different lengths. The set length difference is (a) 0.25 m, (b) 0.35 m, (c) 0.65 m, (d) 0.85 m, (e) 1.35 m, (f) 1.85 m.

We conduct further experimental research to edit the 7-bit Barker phase-coded waveform to endow this waveform with the capability of spectrum sensing. In this experiment, as shown in Fig. 5, the 7-bit Barker phase-coded waveform is sent to DP-MZM1, and the corresponding binary sequence is applied to DP-MZM2. Figure 5(a) shows the temporal waveform of the 7-bit Barker phase-coded waveform from 0 to 1.09375 ns after the signal bandwidth is restricted to 20.7 GHz in the simulation and experiment using a digital low-pass filter. Figure 5(b) is the theoretical temporal waveform of the corresponding binary sequence from 0 to 1.09375 ns. Figure 5(c) shows the temporal waveforms of the binary sequence from 0 to 1.09375 ns after the bandwidth is also restricted to 20.7 GHz in the simulation and experiment. To observe the edited waveform in the experiment, we extract a portion of the optical carrier via an optical coupler and couple it with the edited optical waveform output by DP-MZM2. Then, the coupled optical signal is detected in a PD to convert it into an electrical signal. Subsequently, the electrical waveform is captured using the OSC, and its waveform along with the corresponding simulation results are shown in Fig. 5(d). As can be seen, the two waveforms exhibit a similar variation trend.

Using this edited waveform, the frequency of the SUT can be mapped to the temporal domain. Figure 5(e) shows the generated pulse waveform in a signal period (2 ns) when a single-tone signal at 10 GHz is employed as the SUT. The FWHM of the pulse is 45.8 ps, and the corresponding frequency resolution is approximately 0.84 GHz. Figure 5(f) shows the captured electrical pulses in five individual measurements when single-tone signals at frequencies of 2, 6, 10, 14, and 17 GHz are employed as the SUT. Based on the difference in SUT frequency, pulses appear at different positions in the temporal domain, which correspond to different frequencies. The frequency measurement errors of 400 measurements at the five frequencies are shown in Fig. 5(g). The average measurement values of the five frequencies are 2.0694, 5.9909, 10.0608, 14.0386, and 17.0976 GHz, respectively, while the corresponding standard deviations are 11.5, 49.0, 100.7, 43.8, and 108.9 MHz, respectively. Figure 5(h) shows the time–frequency analysis result of a signal with a sinusoidal time–frequency relationship, a temporal length of 4 µs, and a frequency range from 1 to 17 GHz using the edited waveform. It can be observed that the sinusoidal time–frequency characteristic of the SUT is well restored, which further proves that our proposed waveform editing method can effectively transform a waveform originally lacking frequency measurement and time–frequency analysis capabilities into one with such capabilities.

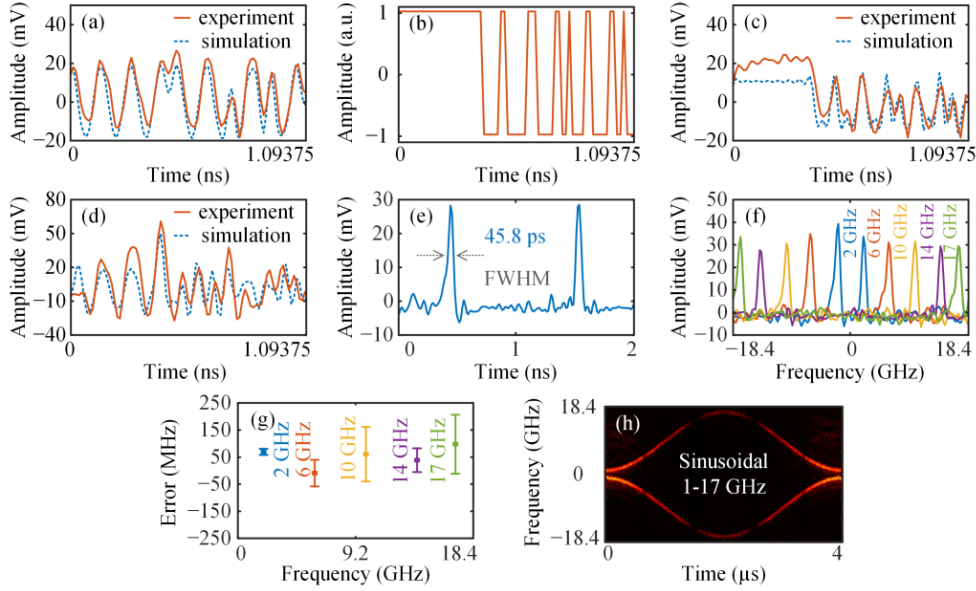

Fig. 5. (a) Temporal waveform of a 7-bit Barker phase-coded waveform in the simulation and experiment. (b) Theoretical temporal waveform of the binary sequence. (c) Temporal waveforms of the binary sequence after the bandwidth is restricted to 20.7 GHz in the simulation and experiment. (d) Temporal waveforms of the edited waveform in the simulation and experiment. (e) The generated electrical pulses in a period when a single-tone signal at 10 GHz is employed as the SUT. (f) The generated electrical pulses when single-tone signals at frequencies of 2, 6, 10, 14, and 17 GHz are employed as the SUT. (g) Frequency measurement errors of the five single-tone signals. (h) Time–frequency diagram of a signal with a sinusoidal time–frequency relationship ranging from 1 to 17 GHz.

3.4 Waveform Editing for Arbitrary Waveforms

To comprehensively investigate and validate the remarkable potential and applicability of the proposed microwave photonic waveform editing method in handling signals in different signal formats, a series of experiments is further carried out.

An LFM waveform ranging from 0 to 10 GHz that does not match the DCF, an NLFM waveform ranging from 0 to 10 GHz, and a waveform with an "E" time–frequency diagram ranging from 1 to 11 GHz are employed, which all have a temporal length of 1.125 ns and cannot be directly compressed into pulses using the DCF (30 dB insertion loss and −6817 ps/nm dispersion) available in our laboratory. These waveforms are respectively sent to DP-MZM1. To enable these waveforms to possess the capability of being compressed into pulses, it is necessary to design special binary sequences based on the accumulation function as given in Eq. (8). Figures. 6(a)–(c) illustrates the temporal waveforms of the three signals (blue dashed line) and the corresponding designed binary sequences (orange solid line) in the simulation. At the end of each waveform period, we add 0.875-ns zeros, thus, the analysis bandwidth is still 36.8 GHz. Figures. 6(d)–(f) are the temporal waveforms of the pulses, respectively compressed from the three edited waveforms in the simulation when no SUT is applied. The FWHMs of the pulses are 42.6, 44.4, and 44.0 ps; the corresponding frequency resolutions are 0.78, 0.82, and 0.81 GHz, respectively. Figures. 6(g)–(i) illustrates the electrical pulses generated within the 2-ns period in the experiment when a 10-GHz single-tone signal is employed as the SUT, and the measurements are conducted using the aforementioned three edited waveforms. The FWHMs of the electrical pulses are 39.7, 43.3, and 42.8 ps, and the corresponding frequency resolution is 0.73, 0.80, and 0.79 GHz, respectively. Theoretically, the width of the pulses generated by the system and the frequency resolution are primarily determined by the temporal length of the waveform (excluding any additionally appended zeros), as indicated by Eq. (10). The minor differences observed in the simulation are primarily caused by the distinct characteristics of each waveform and the precision of the editing process. Figures. 6(j)–(l) illustrates the time–frequency analysis results of an NLFM signal, a dual-chirp LFM signal, and a V-shape LFM signal with a temporal length of 4 µs and a frequency range from 1 to 17 GHz using the three edited waveforms, respectively. It can be observed that the time–frequency characteristics of the three SUTs have been fully restored. Through waveform editing, the three signals, which originally lacked spectrum sensing capabilities, have now acquired the ability to perform time–frequency analysis for wideband signals.

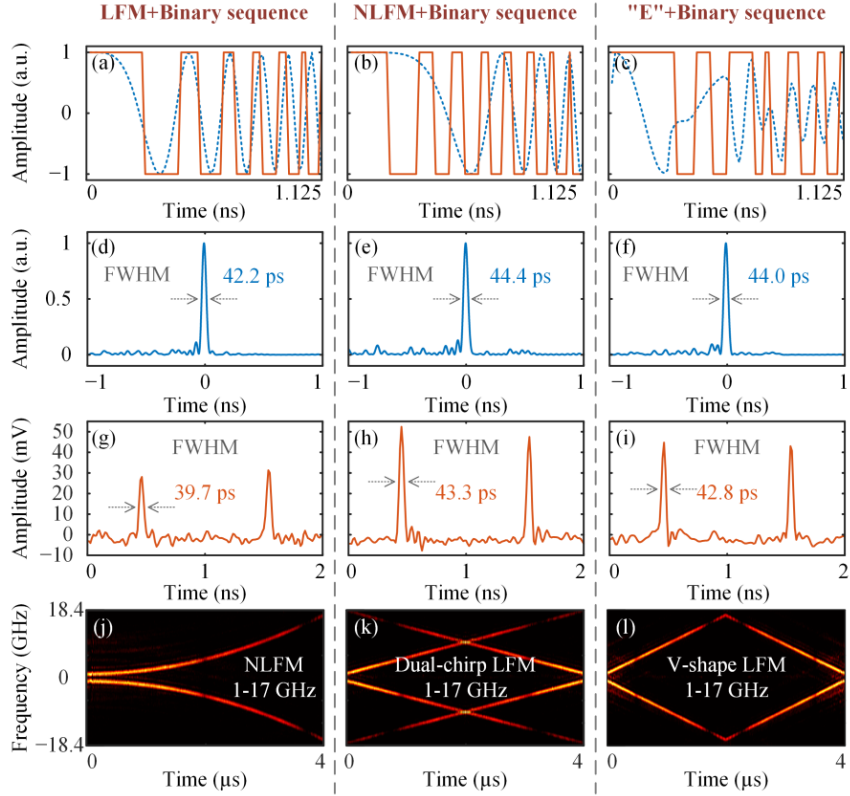

Fig. 6. (a)–(c) Temporal waveforms of the three signals (blue dashed line) and the corresponding binary sequence (orange solid line) in the simulation. (d)–(f) Temporal waveforms of the electrical pulse compressed from the edited waveform in the simulation when no SUT is applied. (g)–(i) Generated electrical pulses in the experiment when a single-tone signal at 10 GHz is employed as the SUT and the measurements are conducted using the aforementioned three edited waveforms. Time–frequency diagrams of (j) an NLFM signal, (k) a dual-chirp LFM signal, and (l) a V-shape LFM signal ranging from 1 to 17 GHz using the three different edited waveforms.

## 4. Discussion

### 4.1 Frequency resolution

As given in Eq. (10), the frequency resolution of this kind of system is only determined by the temporal length of the waveform. It is important to notice that if the analysis bandwidth is expanded by adding zeros at the end of the waveform, the additional length is not included in the aforementioned temporal length used for calculating the theoretical frequency resolution. Then, we validate the relationship between the system's frequency resolution and the temporal length of the waveform by adjusting some parameters derived from the experiments in Section 3.2. The waveform to be edited remains a CW light wave. Unlike the approach taken in Section 3.2, where a 1.125-ns binary sequence was employed for editing, we have reduced the length of the binary sequence to 0.734375 ns in this experiment. Since the total dispersion of the

DCF remains −6817 ps/nm, the 0.734375-ns binary sequence used for waveform editing is identical to the first 0.734375-ns portion of the 1.125-ns binary sequence used in Section 3.2, as shown in Fig. 7(a). Then, the binary sequences are appended with zeros to achieve a temporal length of 2 ns, and the 2-ns signal is periodically repeated. Under these circumstances, the analysis bandwidth is also 36.8 GHz. A single-tone signal at 10 GHz is employed as the SUT. Figure 7(b) shows the electrical pulse waveforms after FTTM using the edited optical waveform when the binary sequence temporal lengths are 0.734375 and 1.125 ns, respectively. As can be seen, the FWHMs of the pulses are approximately 76.4 ps for the case using the 0.734375-ns binary sequence and 46.9 ps for the case using the 1.125-ns binary sequence, corresponding to frequency resolutions of 1.41 and 0.86 GHz. These results are in reasonable agreement with the theoretical frequency resolutions of 1.21 and 0.79 GHz derived from Eq. (10). Therefore, it is verified that the frequency resolution of the system is inversely proportional to the temporal length of the waveform, as given in Eq. (10). Additionally, to validate the time–frequency analysis performance of the proposed system in the case of using the 0.734375-ns binary sequence for waveform editing, an LFM signal ranging from 2 to 17 GHz is employed as the SUT. The time–frequency analysis result is shown in Fig. 7(c). Compared with Fig. 3(f), it can be observed that both results exhibit the same analysis bandwidth, while the frequency resolution in Fig. 7(c) is degraded due to the reduction of the temporal length.

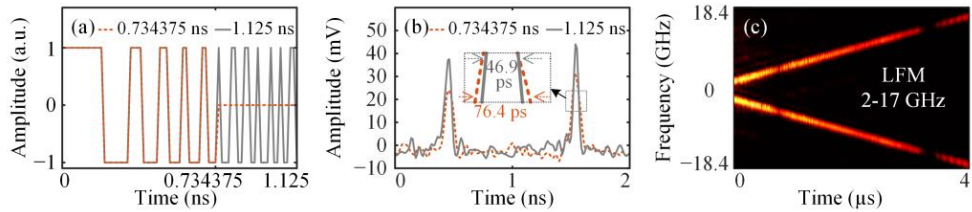

Fig. 7. Temporal waveforms of (a) the binary sequences and (b) the electrical pulse waveforms when employing different binary sequence lengths. (c) Time–frequency diagram of an LFM signal ranging from 2 to 17 GHz when employing the 0.734375-ns binary sequence.

4.2 Dispersion

In the above experiments and discussion, the total dispersion of the DCF is fixed at −6817 ps/nm. When an optical LFM signal is expected to be compressed into an optical pulse, as given in Section 2, the chirp rate of the optical LFM signal needs to match the total dispersion of the DCF, that is $k = -1/(2\pi\beta_2 L)$. Then, through Eq. (9), we find that when only the arbitrary waveform temporal length is considered while disregarding the duty cycle, the system's analysis bandwidth is related to both the temporal length of the arbitrary waveform and the dispersion of the DCF. Theoretically, variations in the total dispersion of the DCF, while achieving the same analysis bandwidth disregarding the duty cycle, affect both the temporal resolution and frequency resolution.

In the experiment detailed in Section 3.2, the total dispersion of the DCF is −6817 ps/nm and the temporal length of the binary sequence is 1.125 ns, and the corresponding analysis bandwidth without considering the duty cycle is 20.7 GHz (−10.35 to 10.35 GHz). This 1.125-ns binary sequence is depicted by the gray solid line in Fig. 8(a). Upon reducing the total dispersion of the DCF to −5396 ps/nm and maintaining an analysis bandwidth of 20.7 GHz without employing the duty-cycle technique, the temporal length of the binary sequence is recalculated to be 0.84375 ns via Eq. (9). This 0.84375-ns binary sequence is designed and illustrated by the red dotted line in Fig. 8(a). Subsequently, the binary sequences are appended with zeros to achieve a temporal length of 2 ns, and the 2-ns signal is periodically repeated. Under these circumstances, the analysis bandwidths considering the duty cycle are 36.8 and 46.4 GHz for the 1.125- and 0.84375-ns binary sequences, respectively. When a single-tone signal at 10 GHz is employed as the SUT, Fig. 8(b) shows the electrical pulse waveforms after FTTM using the edited optical waveform when the dispersion is −5396 and −6817 ps/nm, respectively. It is evident that, when configured with an identical temporal resolution of 2 ns, the system's analysis bandwidth increases as the dispersion decreases because of the different duty cycles. Moreover, despite the FWHM of the pulses being similar in Fig. 8(b), the frequency resolution when the dispersion is lower is worse than that when the dispersion is higher due to the varying analysis bandwidths. The FWHMs of the pulses are approximately 46.1 ps for the case employing a DCF with a total dispersion of −5396 ps/nm and 46.9 ps for the case employing a DCF with a total dispersion of −6817 ps/nm, corresponding to frequency resolutions of 1.07 and 0.86 GHz. Returning to Eq. (10), the $T_r$ values for the two cases are 1.125 and 0.84375 ns, respectively, so it is also evident that the latter case exhibits poorer frequency resolution. The increase in the analysis bandwidth can also be readily observed from the time–frequency analysis results. An LFM signal ranging from 2 to 17 GHz is employed as the SUT when the 0.84375-ns binary sequence is used for waveform editing, and the total dispersion of the DCF is −5396 ps/nm. The time–frequency analysis result is shown in Fig. 8(c). It can be seen that the time–frequency characteristics of the LFM signal are well restored. In addition, due to the introduction of the duty cycle, the final actual analysis bandwidth is 46.4 GHz, which is larger than the previous analysis bandwidth of 36.8 GHz.

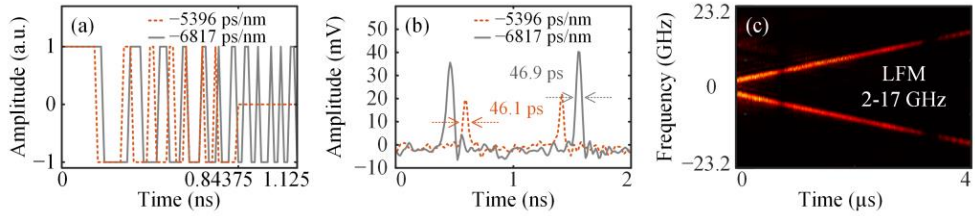

Fig. 8. Temporal waveforms of (a) the binary sequences and (b) the electrical pulse waveforms under different dispersion conditions. (c) Time–frequency diagram of an LFM signal ranging from 2 to 17 GHz when the total dispersion of the DCF is −5396 ps/nm.

4.3 Time delay mismatch

When editing an arbitrary waveform using a binary sequence, precise temporal alignment between the arbitrary waveform and the binary sequence is essential, necessitating careful time-delay matching of the two signals. In the experimental setup, the electrical and optical link delays experienced by the arbitrary waveform and the binary sequence are different. To address this issue, digital-domain time-delay adjustment is employed to synchronize the signal delays accurately. It is crucial to note that mismatched delays can prevent the binary sequence edited arbitrary waveform from being compressed to its narrowest form or even from being compressed into a pulse. This will affect the performance of spectrum sensing, or even make it impossible to perform the spectrum sensing function.

Based on the editing of the LFM waveform with dispersion mismatch in Section 3.4, we further investigate the impact of time delay mismatch on pulse generation by FTTM through adjusting the relative time delay between the binary sequence and the LFM waveform. Figure. 9 shows the temporal pulse waveforms under a single-tone test. As can be seen, as the time delay transitions from matched to increasingly mismatched conditions, the amplitude of the compressed pulses continuously decreases while the FWHM progressively broadens. When the time delay mismatch reaches 1171.875 ns, the temporal pulse completely disappears. Notably, as the delay mismatch further increases to 1406.25, 1640.625, and 1875 ps, the pulses reemerge and gradually narrow again. This phenomenon primarily arises from our use of a 1.125-ns binary sequence to edit the waveform, where the subsequent 0.875-ns segment remains zero, resulting in an overall signal period of 2 ns. When the time delay mismatch further increases to 1406.25, 1640.625, and 1875 ps, the system begins editing the next period of the waveform. Theoretically, a 2-ns delay mismatch should yield results similar to those in Fig. 9(a), as a 2-ns mismatch corresponds to full alignment with the waveform of the next period.

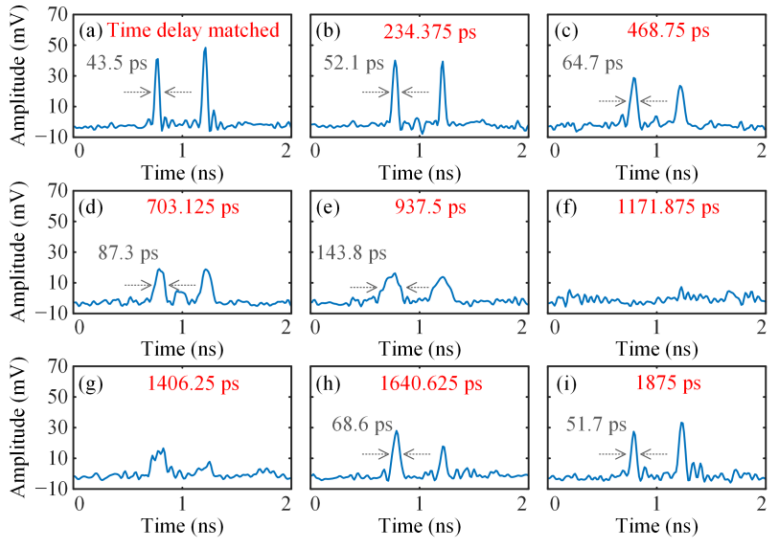

Fig. 9. Temporal waveforms after FTTM under different time delay mismatches between the LFM waveform and the binary sequence.

4.4 Binary sequence bandwidth reduction

In the proposed microwave photonic waveform editing method, arbitrary microwave waveforms can be edited and equipped with the capability to perform spectrum sensing, so radar systems using arbitrary radar waveforms can be upgraded into joint radar and spectrum sensing systems without changing the hardware of radar systems. To achieve the aforementioned objective, in addition to requiring an additional set of microwave photonic structures, a high-speed binary sequence is also needed. Here, we will discuss the relationship among the bandwidth of the original radar waveform, the bandwidth of the binary sequence, and the bandwidth of the edited waveform for spectrum sensing.

As demonstrated in Section 3.2, when compressing a CW light wave with a very small signal bandwidth, the binary sequence requires a bandwidth of approximately 20.7 GHz to achieve a 20.7 GHz analysis bandwidth, as shown in Fig. 3(b). In comparison, in Section 3.4, when the waveform originally has a large bandwidth, for example, the LFM waveform ranging from 0 to 10 GHz, the NLFM waveform ranging from 0 to 10 GHz, and the waveform with an "E" time–frequency diagram ranging from around 1 to 11 GHz, the bandwidth required for the binary sequence is greatly reduced. The spectra of the waveforms before editing, the spectra of the corresponding binary sequences, and the spectra of the waveforms after editing are shown in Fig. 10, where both experimental results and simulation results are given.

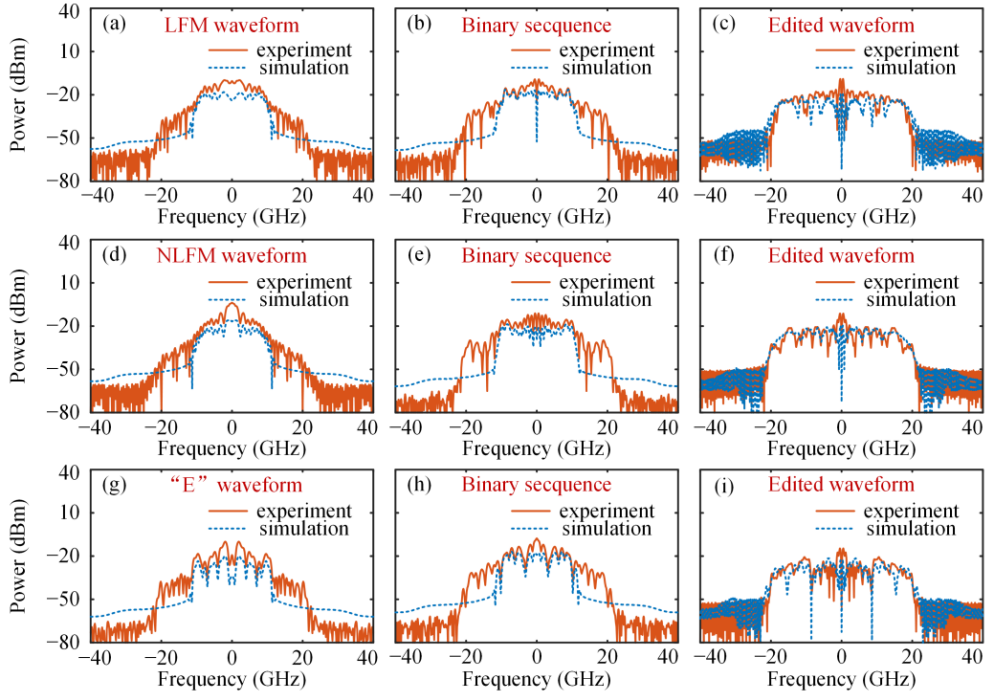

Fig. 10. Spectra of (a), (d), (g) the arbitrary waveforms, (b), (e), (h) the corresponding binary sequences, and (c), (f), (i) the waveforms after editing.

In the experiment, we use a digital filter with a bandwidth of 20.7 GHz to limit the bandwidth of the original waveform and binary sequence. It can be seen that the energy of the original waveform and binary sequence is mainly concentrated within 10 GHz. From the experimental results, it can be seen that the edited waveform occupies a bandwidth of about 20.7 GHz, which directly determines the analysis bandwidth of the system. In the simulation, we directly use an 11-GHz filter to limit the bandwidth of the original waveform and binary sequence. It can be seen that although the original waveforms and binary sequences only have a bandwidth of around 10 GHz, the edited waveforms we obtained still have a bandwidth of around 20 GHz, which is very close to the results in the experiment.

Therefore, by employing the waveform editing method proposed in this work, it is possible to achieve a larger measurement bandwidth using a smaller binary sequence bandwidth while leveraging the bandwidth of the original waveform. Compared with directly utilizing a wideband LFM signal for spectrum sensing, the bandwidth of the binary sequence can be significantly reduced, thereby lowering the difficulty of signal generation. Moreover, when we use binary sequences to edit radar waveforms, the generation of binary sequences only requires a 1-bit DAC. This substantially reduces the generation complexity compared to that of complex LFM signals.

It is important to note that, for the purpose of observing the spectrum of the waveform after editing, we introduce an additional optical carrier in both the simulation and experiment, and beat the optical carrier with the optical sideband after editing. As a result, the generated electrical spectrum not only contains the beat frequency component between the optical carrier and the optical sideband after editing but also includes the self-beat frequency component of the optical sideband. Nevertheless, overall, the simulation results are in agreement with the experimental results. The primary difference between the two mainly arises from the distinct limitations we imposed on the signal bandwidth in the simulation and the experiment.

## 5. Conclusion

In summary, we have proposed and experimentally demonstrated an advanced microwave photonic waveform editing method that enables the editing of arbitrary radar waveforms, equipping them with the capability to perform spectrum sensing. The key contribution of this research lies in the development of a novel waveform editing methodology, which offers a feasible and cost-effective means to expand single-function radar systems into joint radar and spectrum sensing systems. The concept was verified by a series of experiments. Experimental results demonstrated the successful editing of four typical waveforms: a 7-bit Barker phase-coded waveform, an LFM waveform, an NLFM waveform, and a waveform with an "E" time–frequency diagram. These edited waveforms were effectively employed for spectrum sensing across an ultra-wide bandwidth of 36.8 GHz. The system exhibited remarkable performance, characterized by a temporal resolution of 2 ns and a frequency resolution of 0.86 GHz. When compared with directly using an LFM signal for spectrum sensing, the method proposed in this work allows for the attainment of a larger analysis bandwidth by leveraging a binary sequence with a smaller bandwidth in conjunction with the bandwidth of the original radar waveform. Additionally, the binary sequence can be generated using a 1-bit DAC, which is simple and cost-effective. Consequently, this approach can reduce both the signal bandwidth and the generation complexity required for extending a radar system into a joint radar and spectrum sensing system. The method proposed in this work for evolving a radar system into a joint radar and spectrum sensing system opens up new possibilities for transforming single-function systems into multi-functional systems.

**Acknowledgements**
National Natural Science Foundation of China (62371191, 62401207); Space Optoelectronic Measurement and Perception Laboratory, Beijing Institute of Control Engineering (LabSOMP-2023-05); Key Laboratory of Radar Imaging and Microwave Photonics (Nanjing University of Aeronautics and